\documentclass[onecolumn]{aastex631}
\usepackage{CJK}
\usepackage{multirow}

\usepackage{xcolor}

\begin{document}

\begin{CJK*}{UTF8}{gbsn}
\title{Machine Learning Techniques to Distinguish Giant Stars from Dwarf Stars Using Only Photometry - Pushing Redwards}

\author[0000-0001-8470-1725]{Keyi Ding (丁可怿)}
\affiliation{William H.\ Miller III Department of Physics \& Astronomy, Johns Hopkins University, Baltimore, MD 21218, USA}
\affiliation{Department of Astronomy, University of Maryland, College Park, MD 20742, USA}
\email{kyding@umd.edu}
\author[0000-0001-5522-5029]{Carrie Filion} 
\affiliation{William H.\ Miller III Department of Physics \& Astronomy, Johns Hopkins University, Baltimore, MD 21218, USA}
\affiliation{Center for Computational Astrophysics, Flatiron Institute, 162 Fifth Avenue, New York, NY 10010, USA}
\author[0000-0002-4013-1799]{Rosemary F.G.~Wyse}
\affiliation{William H.\ Miller III Department of Physics \& Astronomy, Johns Hopkins University, Baltimore, MD 21218, USA}
\author[0000-0001-6196-5162]{Evan N. Kirby}
\affiliation{Department of Physics and Astronomy, University of Notre Dame, Notre Dame, IN 46556, USA}
\author[0000-0001-8239-4549]{Itsuki Ogami}
\affiliation{National Astronomical Observatory of Japan, 2-21-1 Osawa, Mitaka, Tokyo 181-8588, Japan}
\affiliation{The Institute of Statistical Mathematics, 10-3 Midoricho, Tachikawa, Tokyo 190-8562, Japan}
\author[0000-0002-9053-860X]{Masashi Chiba}
\affiliation{Astronomical Institute, Tohoku University, 6-3 Aoba, Aramaki, Aoba-ku, Sendai, Miyagi 980-8578, Japan}
\author[0000-0002-3852-6329]{Yutaka Komiyama}
\affiliation{Department of Advanced Sciences, Faculty of Science and Engineering, Hosei University, 3-7-2 Kajino-cho, Koganei, Tokyo 184-8584, Japan}
\author{L\'aszl\'o  Dobos}
\affiliation{William H.\ Miller III Department of Physics \& Astronomy, Johns Hopkins University, Baltimore, MD 21218, USA}
\author{Alexander S. Szalay}
\affiliation{William H.\ Miller III Department of Physics \& Astronomy, Johns Hopkins University, Baltimore, MD 21218, USA}


\begin{abstract}
We present our photometric method, which combines Subaru/HSC $NB515$, g, and i band filters to distinguish giant stars in Local Group galaxies from Milky Way dwarf contamination. The $NB515$ filter is a narrow-band filter that covers the MgI+MgH features at $5150$ \AA, and is sensitive to stellar surface gravity. Using synthetic photometry derived from large empirical stellar spectral libraries, we model the $NB515$ filter's sensitivity to stellar atmospheric parameters and chemical abundances. Our results demonstrate that the $NB515$ filter effectively separates dwarfs from giants, even for the reddest and coolest M-type stars. To further enhance this separation, we develop machine learning models that improve the classification on the two-color ($g-i$, $NB515-g$) diagram. We apply these models to photometric data from the Fornax dwarf spheroidal galaxy and two fields of M31, successfully identifying red giant branch stars in these galaxies.

\end{abstract}




\section{Introduction} \label{sec:intro}
Our understanding of the stellar populations within Local Group galaxies has been significantly advanced over the past few decades, largely due to data from wide-field photometric and spectroscopic stellar surveys.
The advent of highly multiplexed fiber-fed spectrographs such as the Prime Focus Spectrograph \citep[PFS;][]{sug14, tak14} on the Subaru Telescope or the Dark Energy Spectroscopic Instrument \citep[DESI;][]{desi22} on the Mayall Telescope has further accelerated the study of Local Group galaxies through analyses of the chemodynamics of their member stars \citep[e.g.,][]{dey23}.  Careful target selection is critical to the overall efficiency and success of spectroscopic surveys, and surveys of Local Group galaxies in particular benefit greatly from the identification of likely member stars, as opposed to stars from the Milky Way that lie along the line-of-sight,  prior to the assignment of fibers. Generally, the foreground Milky Way stars are main sequence (dwarf) stars while the desired targets are evolved red giant stars; the different distances and intrinsic luminosities of these two populations conspire to place them in similar loci in the observed broad-band color-magnitude diagram, so that straightforward cuts in this plane, such as isolating sources that lie close to red giant branch (RGB) isochrones, are not sufficient to distinguish members from non-members. The addition of photometry in a narrow-band filter that is designed to encompass a suitable gravity-sensitive absorption feature can provide the extra dimension to break this degeneracy.

The use of photometry obtained from filters of various widths to produce estimates of stellar atmospheric parameters such as metallicity and gravity is long-established \citep[see the review by][]{stromgren66}. The Str\"omgren system is well-suited for stars of F/G spectral type; the analysis of cooler, later-type stars benefits from the addition of a medium/narrow-band filter that encompasses the gravity-sensitive magnesium features around 5100\AA\ \citep[see Table~3 of][]{pritchet1977}.  The utility of such a filter in removing foreground dwarf stars has been demonstrated in many analyses of (metal-poor) red giants in globular clusters \citep[perhaps most famously by][]{SZ78} and in the distant field halo \citep[e.g.,][]{maj00}. The strength of the magnesium absorption features also depends on the metallicity/magnesium content of a star, and, as discussed by \cite{guh03} in the context of use of the DDO51 filter \citep[defined by][]{cla79} to select spectroscopic targets in the outer regions of M31, there may be a resulting bias against more metal-rich RGB stars, as they appear more similar to dwarf stars in terms of the strength of the absorption. \cite{guh03} found no obvious bias, but for metallicities above those probed by their fields (mean photometric metallicity of $\sim -0.5$~dex or below) the concern remains. Indeed, comprehensive wide-area spectroscopic surveys of M31 that encompass the metal-rich disk(s) and use narrow-band filters to inform target selection, such as planned by the PFS collaboration,  need to take particular care to optimize the separation of foreground, low-mass M dwarfs from cool, metal-rich M giants in M31.  

To address this point, this paper investigates the effectiveness of the narrow-band filter $NB515$ designed for use with  Hyper Suprime-Cam (HSC) on the Subaru Telescope in distinguishing likely foreground contamination and member red giant stars of target Local Group galaxies, across the full range of metallicity, into the M-star regime. The $NB515$\/ filter has a bell-shaped transmission curve centered at 5150\AA\ with a FWHM of 77\AA, encompassing the  Mgb stellar absorption feature.
It was first used by \cite{kom18} in a photometric study of the outer halo and NW  stellar stream of M31: a polygon was drawn on the reddening-corrected HSC (($NB515-g$), ($g-i$)) two-color diagram to select likely giants in the M31 halo for further analysis. Again, RGB stars in these fields are expected to be predominantly metal-poor. A more statistically motivated method to select likely M31 giants using $NB515$ data was developed by \cite{oga24}, whereby the associated uncertainties in the photometry were incorporated to determine a statistical criterion that separates the dwarfs and giants in the two-color space. That paper focused on previously identified substructure in the stellar halo of M31, and limited the analysis to stars with ($g-i$)$_0$ bluer than 2.5 (see their Figure~7, lower rightmost panel).

This study investigates extending the use of $NB515$ into the M-star spectral class, so that unbiased chemical profiles of the target galaxies can be obtained.  We adopt a data-driven approach and model the sensitivity of the $NB515$ filter to stellar atmospheric parameters and chemical abundances through the creation of synthetic photometry for stars of a wide range of spectral types, based on large public empirical stellar spectral libraries. We first demonstrate that the dwarfs and giants are separated in the two-color space even for M stars, although the separation decreases as temperature and/or metallicity increases. We then employ machine learning methods to conduct dwarf-giant separation with $NB515$ data. 

The paper is organized as follows: Section~\ref{sec:data} describes our method for constructing the synthetic photometry catalog from empirical spectral libraries and the analysis of the distribution of various stellar spectral types in the two-color diagram. We detail our method to develop machine-learning models for target selection in Section~\ref{sec:applications}, with our results  for fields in two Local group galaxies (M31 and the Fornax dwarf spheroidal galaxy) being presented in Section~\ref{sec:discussion}. Our conclusions are summarized in Section~\ref{sec:conclusion}.

\section{Synthetic Photometry of Galactic Stars} \label{sec:data}
\subsection{Synthetic Photometry Calculation}

The calculation of synthetic colors, namely ($g-i$) and ($NB515-g$), is the first step in our investigation of the expected distributions of dwarfs and giants on the ($g-i$, $NB515-g$) two-color diagram. The spectral features of M-stars are particularly challenging to model in synthetic spectral libraries,  \citep{jon96, pas16, 2024ARA&A..62..593H}, due to the presence of complex and often poorly known molecular lines at these low effective temperatures,   and thus we chose to use empirical spectral libraries for more straightforward comparisons with observational data. 

\eject
\subsubsection{Spectral Libraries}
We utilized both the MaNGA Stellar Library (MaStar, \citealt{yan19}) and the X-shooter Spectral Library DR3 \citep[XSL,][]{chen14, gon20, ver22} with the MaStar Spectral Library as our primary source,  as the spectra it contains are both fluxed and cover a wide wavelength range ($3622-10354~\text{\AA}$), including  the $g$ and $i$ filters. The MaStar library also contains a sufficiently large  sample of stars with  comprehensive coverage of stellar parameter values, which is especially beneficial for modeling the stellar population in both the Milky Way and external galaxies. The sample we selected for analysis consists of 31577 high-quality spectra from 14514 unique stars in the MaStar library. 

Additionally, where necessary, we supplemented the MaStar catalog using the XSL, which contains  830 spectra of 683 stars, with moderate to high resolution and near-ultraviolet-to-near-infrared  wavelength coverage ($3500-24800~\text{\AA}$). Specifically, XSL includes a sample of asymptotic giant branch (AGB) stars and post-AGB stars, augmenting the coverage of redder stars in  MaStar. The XSL Data Release 3 also adds spectra of 20 archival M-dwarf stars, which is particularly helpful for our goal of finding the separation between M-dwarfs and M-giants on the two-color diagram.

It should be noted that  we assume perfect flux calibration for both spectral libraries in the following analysis, while acknowledging the differences in data acquisition: MaStar utilizes a fiber-fed system with spectra collected by the BOSS spectrograph \citep{sme13} and the MaNGA fiber bundles \citep{dro15}, whereas XSL was obtained  using the slit-based X-shooter spectrograph on the ESO Very Large Telescope \citep{ver11}. The descriptions of how the fluxing was performed for MaStar and XSL, respectively, can be found in \cite{yan16, yan19} and \cite{chen14}.

\subsubsection{Extinction Correction}
Given that our method uses empirical spectra as the input for the derivation of the synthetic photometry, we must first apply an extinction correction to the spectra before we can create properly dereddened synthetic colors. 

In XSL, 748 out of 830 spectra have been corrected for Galactic dust extinction provided by the XSL team \citep{chen14}; thus in our calculation, we filtered out spectra without extinction correction. On the other hand, none of the spectra in MaStar were corrected for extinction, and we adopted the following method to correct them. The MaStar Cross-match Catalog provides $E(B-V)$ reddening for each star given by the Bayestar19 3D dustmap \citep{green19} using the median of the photogeometric distance posterior from Gaia EDR3 \citep{2021AJ....161..147B}. We converted the $E(B-V)$ reddening to extinction in the Pan-STARRS 1  $g$ band using the Bayestar19 extinction coefficient $R_g = 3.518$ from \citep{green19}. To match the monochromatic extinction curve that \cite{green19} used, we took the \cite{sch16} extinction curve and calculated the normalization coefficient, which scales the $g$ band extinction in \cite{sch16} to match the $g$ band extinction of stars from \cite{green19}. As described below, we use the scaled extinction curve to correct each spectrum for extinction and to calculate the synthetic colors of MaStar spectra both with and without extinction correction.

\subsubsection{Synthetic Colors}
\begin{figure*}[ht]
                \centering
                 \includegraphics[width=0.95\linewidth]{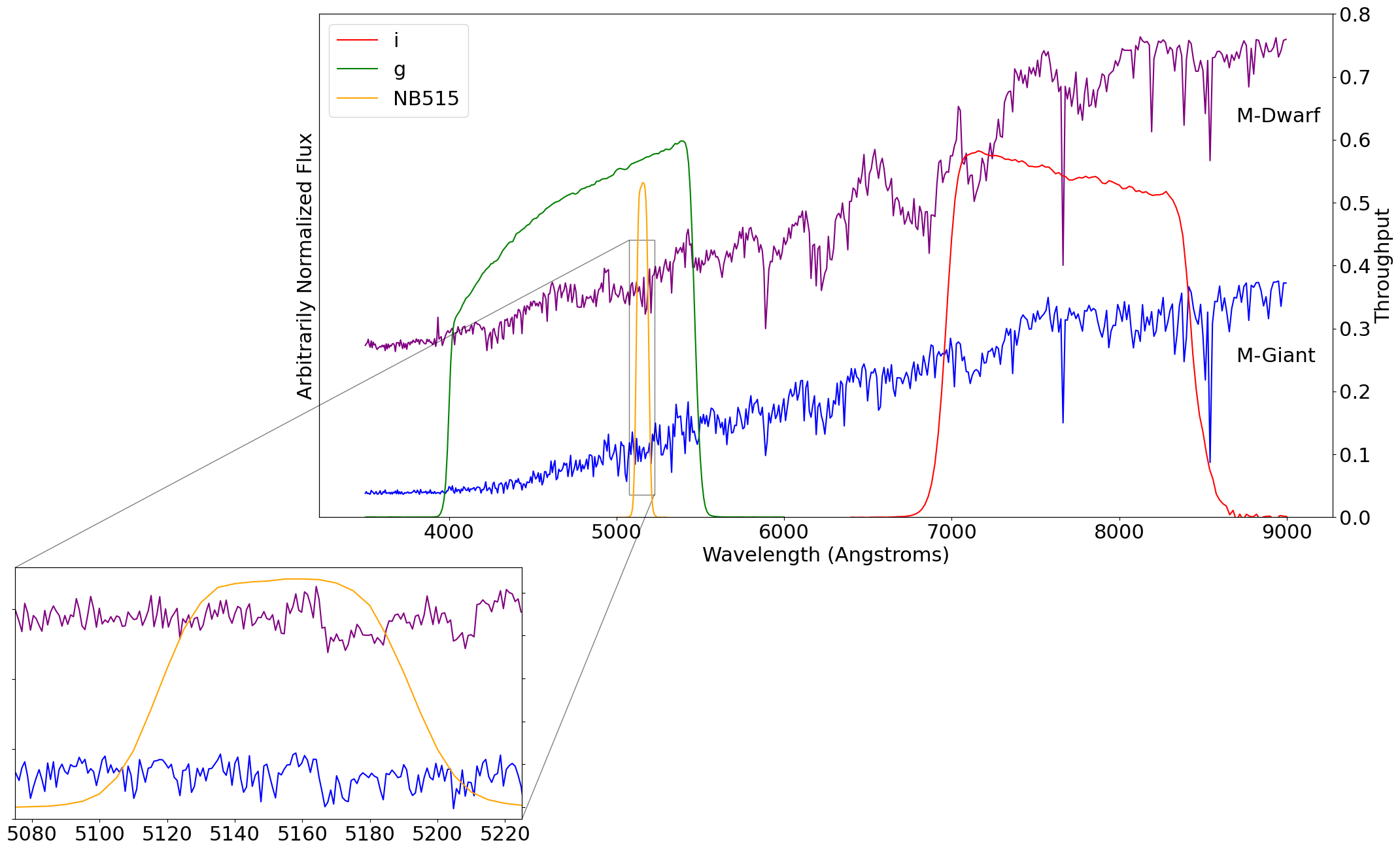}
\caption{Total response curves of the  $g$, $i$, and $NB515$ filters of HSC, overlaid by the observed spectrum of an M-giant (blue) and of an M-dwarf (purple) from the X-shooter Spectral Library. Both spectra have been down-sampled to a resolving power of 2000 and arbitrarily normalized for clarity.  The zoomed-in plot highlights the $NB515$ filter, which is centered around $5150\text{\AA}$ and includes the MgH+Mgb absorption features.
\label{fig:filters}}
\end{figure*}

After the extinction correction, we calculated synthetic $g-i$ and $NB515-g$ colors by convolving the photometric filter curves over the spectra following the derivation in \citep{cas14}. Figure \ref{fig:filters} displays the total response curves of the HSC $g$, $i$, and $NB515$ filters, overlaid with the spectra of an M-giant and an M-dwarf from the X-shooter Spectral Library. While calculating synthetic magnitudes requires information on the stellar radius, we avoided such a requirement and dependence on stellar models by calculating only the synthetic colors. In the calculation, we filter for non-negative flux within the wavelength range of each filter curve to avoid numerical errors and interpolate the filter curve using \texttt{np.interp} function such that the filter data is on the same scale as spectral data. We then numerically integrate the flux in each bandpass using \texttt{scipy.integrate.simps} and convert the integrated flux to AB magnitudes following equation 6 in \cite{cas14}. We compute and convert the integrated flux in each filter, then subtract the value of one filter from another to derive the synthetic color.

\subsubsection{Crossmatching with Photometry Surveys}
We crossmatch our synthetic photometry catalog (without extinction correction) with all-sky photometric surveys to assess the quality of our calculated synthetic colors and obtain apparent magnitude data for subsequent machine-learning models. The MaStar sample is selected from multiple photometric surveys including SDSS \citep{bla17, york00, gun06, gun98, doi10, fuk96, adb22},  PanSTARRS-1 \citep{cha16, mag20,mag20-2, mag20-3,  wat20, fle20}, and APASS \citep{hen15}, and the MaStar Cross-match Catalog provided by the MaStar team includes $g$ and $i$ magnitude data from these surveys. For XSL, we conduct a 1~arcsecond crossmatch with PanSTARRS-1 and SkyMapper DR4 \citep{chr24} using TOPCAT \citep{tay11}. If a star has multiple matches across different surveys, we selected the one with the $g-i$ color that most closely matched our synthetic $g-i$ color (despite minor differences in filter designs). We also reject crossmatched observational data if the difference between observed and synthetic $g-i$ colors does not fall within the range of three times the crossmatched observational photometric uncertainties. A more detailed discussion of the resulting crossmatch is presented in Appendix \ref{app:cm}.

\subsection{Subaru HSC Observational Data}
Based on the synthetic photometry calculated using MaStar and XSL spectra, we developed a machine learning method to distinguish dwarfs and giants. We applied the method to the Fornax dSph and two selected fields in M31. Fornax was observed using Subaru/HSC for the purpose of pre-observation target selection for the Galactic Archaeology survey within the Subaru Prime Focus Spectrograph (PFS) Subaru Strategic Program (SSP)\@. For M31, we analyze the same data used by \cite{oga24}, which combines HSC $NB515$ data of M31 following the strategy and reduction of \cite{kom18} with $g$ and $i$ data from the Pan-Andromeda Archaeological Survey (PAndAS, \citealt{mcc18}). We specifically analyzed 2 fields of M31, with one at the inner halo (PFS\_FIELD\_6 in Table 1 of \cite{oga24}, $\alpha(J2000)=00^h41^m42''.9$, $\delta(J2000)=+42^ \circ 54'25''.0$) and one at the northwest (NW) stream (M31\_009 in Table 1 of \cite{oga24}, $\alpha(J2000)=00^h10^m24''.5$, $\delta(J2000)=+46^ \circ 49'07''.0$). In our analysis, we limit stars in M31 down to $i_0 = 24$, while a magnitude cut at $g=24$ was already applied to the Fornax catalog of photometry.

\eject
\subsection{Two-color Diagram}
\begin{figure*}[ht]
                \centering
                 \includegraphics[width=0.95\linewidth]{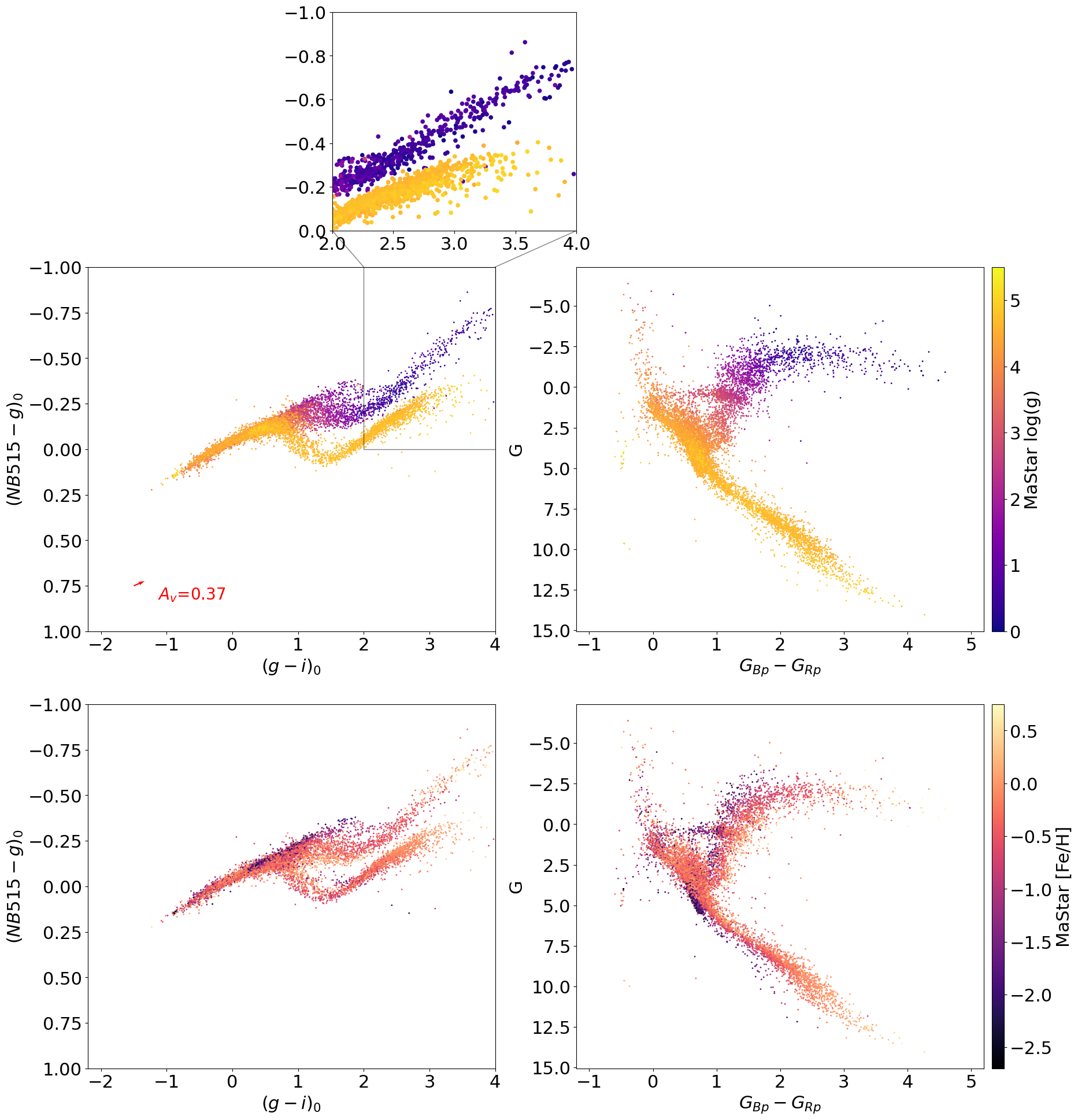}
\caption{$Gaia$ color-magnitude diagram of stars in the MaStar Library and two-color diagram of MaStar synthetic photometry, color-coded by $\log(g)$ and [Fe/H] from the mean MaStar Library parameters. A reddening vector is shown in the top-left panel, with length corresponding to the mean extinction of the catalog ($A_v = 0.37$).
\label{fig:synphot}}
\end{figure*}

\begin{figure*}[ht]
                \centering
                 \includegraphics[width=0.95\linewidth]{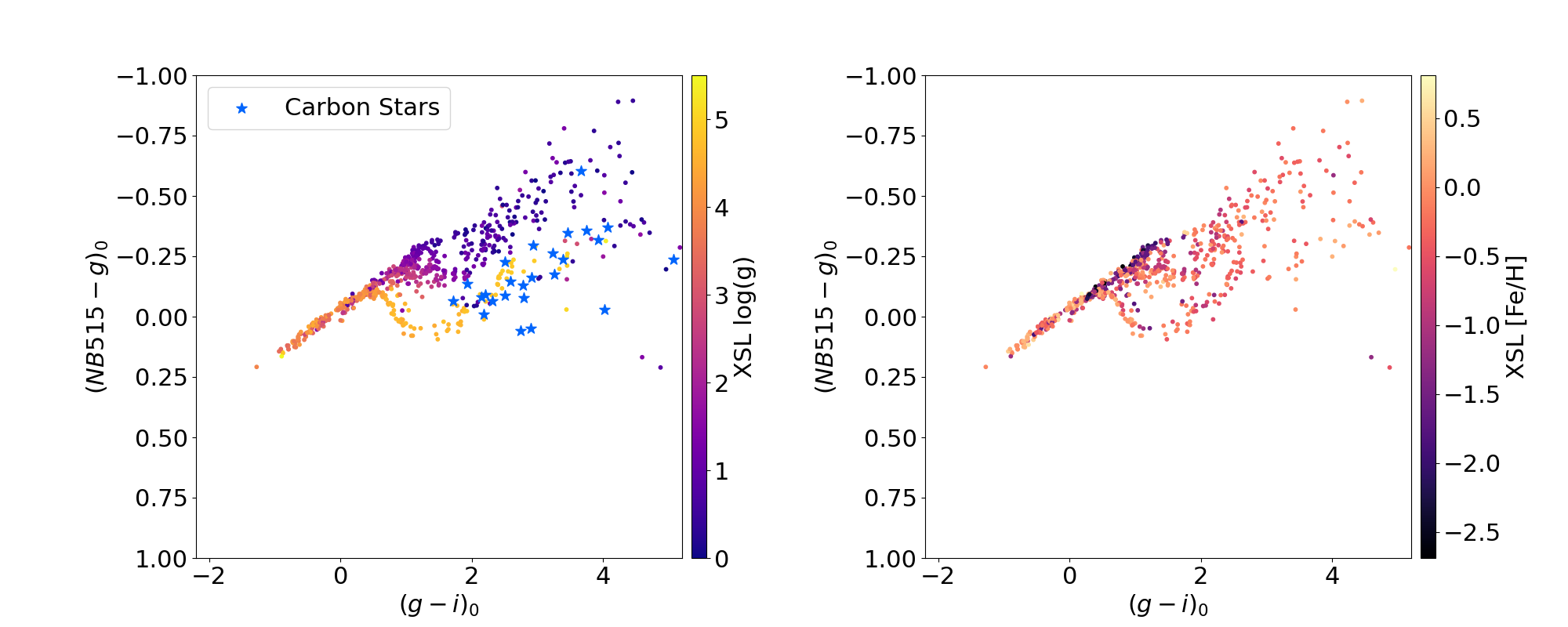}
\caption{Two-color diagrams of XSL synthetic photometry, color-coded by $\log(g)$ and [Fe/H] from the XSL parameters.
\label{fig:xshooter}}
\end{figure*}

To investigate $NB515$'s sensitivity to stellar parameters and chemical abundances, we made two-color diagrams of our derived MaStar synthetic photometry catalog and color-coded by parameter estimates provided in the MaStar Library \citep{2020ApJ...899...62C, 2022A&A...668A..21L, 2022MNRAS.509.4308H, 2022AJ....163...56I} and XSL \citep{ver22}. The MaStar catalog provides parameters from 4 independent pipelines and their unweighted average, and we use these average parameter estimates from the four pipelines in our analysis. The top two panels of Figure \ref{fig:synphot} present the $Gaia$ color-magnitude diagram \citep{gaia16, gaia23} of stars in the MaStar Library and the synthetic two-color diagram color-coded by $\log(g)$ from the MaStar Library parameters. The absolute, extinction-corrected magnitudes are obtained by shifting the apparent $Gaia$  $G$ magnitudes to the \texttt{r\_med\_photgeo} distances estimated in $Gaia$ EDR3 \citep{bai21}. In Figure \ref{fig:xshooter}, we color-code the synthetic colors of stars in XSL with $\log(g)$ from the XSL pipeline. Both figures indicate that the high surface gravity dwarfs and low surface gravity giants are well-separated on the two-color space when $1<(g-i)_0<2.5$. The lower '\checkmark'-shape sequence is composed of dwarf stars with $\log(g)>4$, and the giants are scattered on the upper part of the two-color diagram. Our subsequent analysis primarily focuses on cool M-stars with $(g-i)_0$ colors redder than 2.5, which have traditionally been excluded from analysis (see e.g., \citealt{oga24}) due to the less prominent distinction between dwarfs and giants at these redder colors. The positions of stars on the two-color diagram align closely with the more widely used $[M-T_2, M-DDO51]$ two-color diagram introduced by \cite{2000AJ....120.2550M}. A detailed comparison of the colors can be found in Appendix \ref{app:ddo51}.

We identified a subset of known carbon stars in XSL that lie along the typical dwarf sequence and have red colors $(g-i)_0 > 2$ (Figure~\ref{fig:xshooter}). For these stars, the strong carbon $C_2$ bandheads within the NB515 filter make it difficult to detect the gravity-sensitive magnesium absorption feature used to distinguish dwarfs from giants \citep[see Figure~1 of][]{gre13}. Although our method does not apply to carbon stars, their general rarity in Local Group galaxies means this limitation has a negligible impact on our subsequent analysis.

In addition to the sensitivity of the $NB515-g$ color to surface gravity, we also detected a secondary sensitivity to [Fe/H] (see also \citealt{hong25}). The bottom two panels of Figure \ref{fig:synphot} present the $Gaia$ color-magnitude diagram of stars in the MaStar Library and the synthetic two-color diagram color-coded by [Fe/H] from the MaStar Library parameters, and the right panel of Figure \ref{fig:xshooter} shows the two-color diagram of stars in XSL with [Fe/H] from the XSL pipeline. Along the sequence of the giant stars, there is a gradient of [Fe/H] along the $NB515-g$ color with the most metal-poor giants generally having the more negative $NB515-g$ color index. The finding on $NB515-g$'s sensitivity to [Fe/H] is consistent with Figure 6 of \cite{kom18}, who used empirical and template stellar spectra from \citet{1998PASP..110..863P} and ATLAS9 \citep{2003IAUS..210P.A20C} catalogs to explore the position of dwarfs and giants of various [Fe/H] on the two-color diagram.

Since the NB515 filter is centered on the MgH+Mgb absorption features, the synthetic $(NB515-g)_0$ color is also sensitive to stellar alpha-elements abundances (see Appendix \ref{app:alpha}). A more detailed analysis of the NB515 filter's sensitivity to chemical abundances is presented by \cite{hong25}.

\section{Application} \label{sec:applications}
Using synthetic photometry, we demonstrate that dwarfs and giants can be effectively separated on a two-color diagram, making this technique valuable for dwarf/giant classification and target selection in studies of nearby galaxies. However, as illustrated in Figure \ref{fig:synphot} and Figure \ref{fig:xshooter}, the distinction between dwarfs and giants becomes less prominent for redder stars when $(g-i)_0 > 2.5$. Additionally, photometric uncertainties can further shift the colors on the diagram, blurring the separation between dwarfs and giants. To address this issue, we train machine learning models to identify the underlying decision boundaries that distinguish likely dwarfs from giants.

\subsection{Stellar populations modeling} \label{sec:population}
Since the machine learning models learn the underlying decision boundaries from the training data, it is important to construct training sets with our derived MaStar and XSL synthetic photometry that represent the expected stellar populations in the observations. This is done by modeling the expected stellar populations in both foreground and target galaxies, utilizing a combination of theoretical models and existing spectroscopic observations.

\subsubsection{The Fornax  dwarf spheroidal galaxy}

 The Fornax dwarf galaxy, a satellite of the Milky Way, consists of three distinct stellar population groups --- old, intermediate-aged, and young \citep{2012A&A...544A..73D, lem14}. These populations display a diverse range of metallicities, with a noteworthy radial-metallicity gradient: the central region of Fornax predominantly hosts stars richer in metals compared to the outer regions \citep{2012A&A...544A..73D}. This mix of populations makes Fornax an excellent candidate for testing our methodology across various stellar types.

To construct representative training sets for the stellar populations in Fornax, we sample giant stars from our synthetic photometry catalog. These stars have metallicities matching the spectroscopic metallicity distribution function (MDF) from \cite{2010ApJS..191..352K}, with $\text{[Fe/H]}=-1.05_{-0.25}^{+0.30}$, and $(g-i)_0$ colors that encompass the observed color range for giants in Fornax.

It is worth noting that satellite galaxies of the Milky Way, such as Fornax, are typically metal-poor, with the consequence that its member stars do not extend to the reddest, coolest M-giants. As shown in Figure \ref{fig:synphot}, the distinction between M-dwarfs and M-giants is less pronounced than for hotter stars. To evaluate the performance of our method on M-stars, we need a more metal-rich population, such as that found in the inner halo of M31.  

\subsubsection{M31}
We selected two HSC fields of view in M31's stellar halo: one within the NW stream and another in the inner halo. The inner halo hosts more metal-rich, redder stellar populations \citep{esc20}, whereas the NW stream, with its lower metallicity and surface brightness \citep{gil14, woj23}, likely contains fewer M31 member giants. As a result, we create separate training sets for the two M31 fields.

Considering the gradient of Milky Way stars from the northwest to southeast of M31, arising from the orientation of the Milky Way disk along the direction of M31, the northwest fields are expected to adequately represent a field dominated by Milky Way dwarfs for testing our machine-learning accuracy. We approximate the MDFs of the fields in the inner halo of M31 using skew-normal distributions based on \cite{gil14} with a mean of $-0.2$, width of 1, and asymmetry parameter of $-5$. For the NW field, we adopt a photometric metallicity of $[Fe/H]=-1.3^{+0.05}_{-0.04}$ from \citep{oga24} and apply a skewed Gaussian MDF\@. We then sample giant stars for each field from our catalog, verifying that their metallicities lie within these respective distributions.

\subsubsection{Foreground Milky Way Stars}
The foreground population is modeled using the Besan\c{c}on model of stellar population synthesis of the Galaxy \citep{2012ASSP...26..171R, 2014A&A...564A.102C} along the line of sight to the target galaxies. The Besan\c{c}on model  adopts theoretical distribution functions to simulate Galactic stellar populations within a specified field of view that are directly comparable to survey observations.

In the Besan\c{c}on simulation, we define a square field centered on the target galaxy with a size matching the real area of the HSC observations of each target galaxy, and a progressive distance interval from 0 kpc to 100 kpc. We adopt the SDSS $ugriz$ plus $JHK$ photometric system and apply a magnitude cut in the $i$ band from 15.6 to 24 and a color cut in $(g-i)_0$ from 0 to 5 to match the HSC observations. The model simulates a catalog of artificial stars with photometry (with extinctions) and associated uncertainties in each band, along with stellar parameters such as $T_{eff}$, $\log(g)$, [Fe/H], mass, and radius. To refine the sample, we applied a surface gravity cut at $\log(g)>3.5$ to restrict the sample to dwarf stars, and another cut at $\log(g)>6$ to eliminate potential white dwarfs. 

To match each entry in the Besan\c{c}on simulation to stars in the MaStar synthetic photometry catalog, we use the normalized Euclidean distance of [$T_{eff}$, $\log(g)$, [Fe/H], $(g-i)_0$] as matching metrics. For each Besan\c{c}on simulated star, the two closest dwarf stars with the smallest standardized Euclidean distances in the MaStar catalog are identified. This approach effectively doubles the dataset size, enhancing the robustness of our machine-learning model. Additionally, we assign the line-of-sight distances from the Besan\c{c}on simulation to the selected best-matched stars, which are later used to estimate photometric uncertainties. Finally, this matching process provides a best-matched estimate of the $NB515-g$ color for each star in the foreground population.

Along the line-of-sight to M31, the foreground contamination is primarily dominated by the Milky Way's thick disk and a secondary source from the halo. The thick disk is characterized by the intermediate-metallicity population with $\text{[Fe/H]} \approx -0.5$, while the halo contains a metal-poor population with $\text{[Fe/H]} < -1$. On the other hand, the Milky Way thick disk is slightly less prominent in the foreground of Fornax due to the higher galactic latitude, but still is the major source of contamination along with the Galactic halo. The combination of the Milky Way thick disk and halo as the major source of contamination leads to an asymmetric [Fe/H] with a peak at $-0.5$ reflecting the thick disk, plus a metal-poor tail extending to $-3$ coming from the halo. The same effect can also be seen in our sampled foreground stars as a higher concentration of thick disk stars at the color range $2 < (g-i)_0 < 3$, and more scattered halo stars at $(g-i)_0<1.5$.

\subsubsection{Addition of  Uncertainties to the Artificial Photometry}

After selecting stars from our synthetic photometry catalog to produce mock foreground and member populations, we introduce artificial photometric uncertainties into the training set to simulate the observed color-magnitude and two-color diagrams. The process involves estimating the expected apparent magnitudes for each entry in the training set. For member stars, we shift the observed apparent $g_0$ and $i_0$ magnitudes to the distance of the target galaxies, assuming a uniform distance for all members. For foreground dwarfs, the observed $g_0$ and $i_0$ magnitudes are shifted to the simulated line-of-sight distances from the Besan\c{c}on model. We next fit three univariate spline curves using the \texttt{scipy.interpolate.UnivariateSpline} function to estimate appropriate photometric uncertainties. The first curve models  uncertainties ($g_{err}$) as a function of the observed $g_0$ magnitude from the HSC data of the target galaxies. The second curve models  uncertainties ($NB515_{err}$) as a function of the observed $g_0$ magnitude. It is important to highlight that $NB515$ magnitudes are neither calculated nor available in our observed photometric catalog while we have collected observed $g_0$ and $i_0$ magnitudes; therefore, we derive the $NB515_{err}$ uncertainties using the fitted curve based on $g_0$ magnitudes. The third curve models  uncertainties ($i_{err}$) as a function of the observed $i_0$ magnitude. 

To expand the training set and enhance the robustness of the machine-learning model, we generate 3 new pairs of [$(g-i)_0$ and $(NB515-g)_0$] colors for each entry. These pairs are sampled from a multivariate-normal distribution centered on the original synthetic $(g-i)_0$ and $(NB515-g)_0$ colors, using a covariance matrix derived from the previously estimated uncertainties (assuming no covariance between the colors and adding the errors in quadrature). Additionally, three new $g$ magnitudes are sampled from a Gaussian distribution centered on the shifted $g$ magnitude, with the width determined by the corresponding estimated $g$~magnitude uncertainty. This sampling process ensures a more comprehensive representation of the synthetic stellar populations, contributing to more effective training of the machine-learning algorithm.

\begin{figure*}[ht]
                \centering
                 \includegraphics[width=0.95\linewidth]{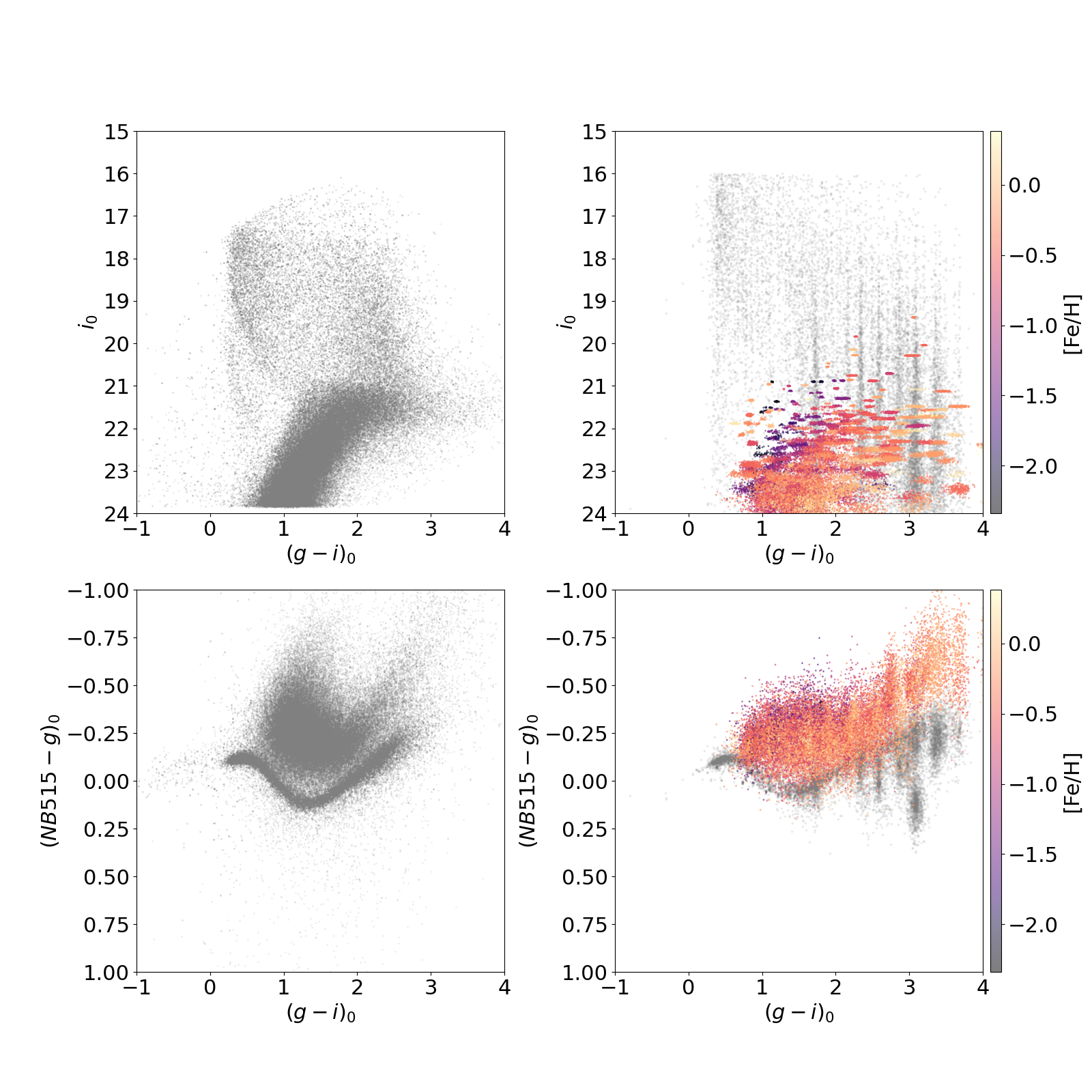}
\caption{Color-magnitude and two-color diagrams of the observed data (left) and training set (right) for this field in the inner halo of M31, with giants in the training set color-coded by [Fe/H].
\label{fig:training}}
\end{figure*}

Figure \ref{fig:training} presents the color-magnitude and two-color diagrams of the observed data (left) and the training set (right) for this field in the inner halo of M31, with the giants in the training set being color-coded by [Fe/H]\@. While the training set resembles the observed data to some extent, certain caveats in our modeling process need to be considered. A discrete pattern in the color of the foreground population potentially arises from the analytical assumptions in the synthetic Galactic model calculations. Additionally, the empirical stellar libraries do not uniformly cover the parameter space of the M31 populations, leading to multiple data points being drawn from the same entry in the synthetic photometry catalog. This results in the \lq clumps' observed in the CMD of the giants. These caveats might contribute to the limitations discussed in Section~\ref{sec:limitation}.

\subsection{Machine Learning}
In our setup, the model takes the $(g-i)_0$, $(NB515-g)_0$ colors, and $g_0$ magnitude as input and predicts normalized membership-probability vectors. The training data are labeled with $[P_{member}, P_{nonmember}]$ probability vectors, such that all member stars in the training set have the label [1, 0] and nonmember stars have the label [0, 1], and the machine learning model aims to predict similar probability vectors. The neural network contains five linear layers with Sigmoid activation functions between layers and a Softmax function as the output to normalize the predicted probability. The Sigmoid function is a bounded, differentiable, `S-shaped' function that introduces non-linearity to the decision boundaries, and thus the boundaries will be curves rather than straight lines. The model is trained using Binary Cross-entropy loss as the objective function to measure the similarity between the probability vector predicted by the machine learning model, $p(\hat{z})$, and the true label probability vector, $q(\hat{z})$. The Binary Cross-entropy loss is measured as a number between 0 and 1, with 0 being a perfect model and the machine-learning model aims to minimize the loss. We optimize the model with a root mean square propagation (RMSprop) optimizer \citep{tieleman2012rmsprop}, which is an extension of the Stochastic Gradient Descent algorithm that takes adaptive learning rates and ensures a faster convergence to the minima. It typically takes 120 epochs for the model to converge without significant overfitting. The architecture and mathematical basis of the training method are detailed in Appendix~\ref{app:ml}.

\eject
\section{Results and Discussion} \label{sec:discussion}

\subsection{Fornax}
\begin{figure*}[ht]
                \centering
                 \includegraphics[width=0.95\linewidth]{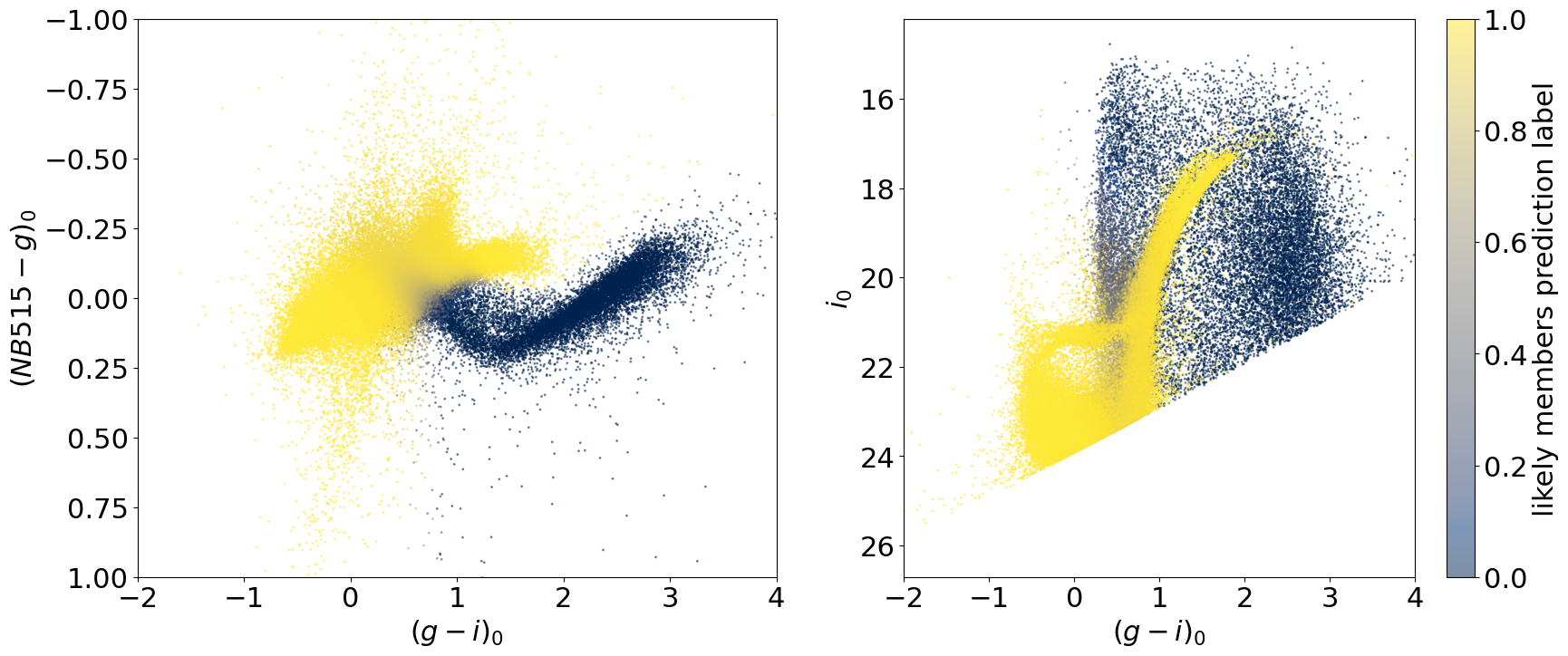}
\caption{Two-color diagram and color-magnitude diagram of Fornax, color-coded by the predicted membership label.
\label{fig:fornax}}
\end{figure*}

\begin{figure*}[ht]
  \centering
  \includegraphics[width=\textwidth]{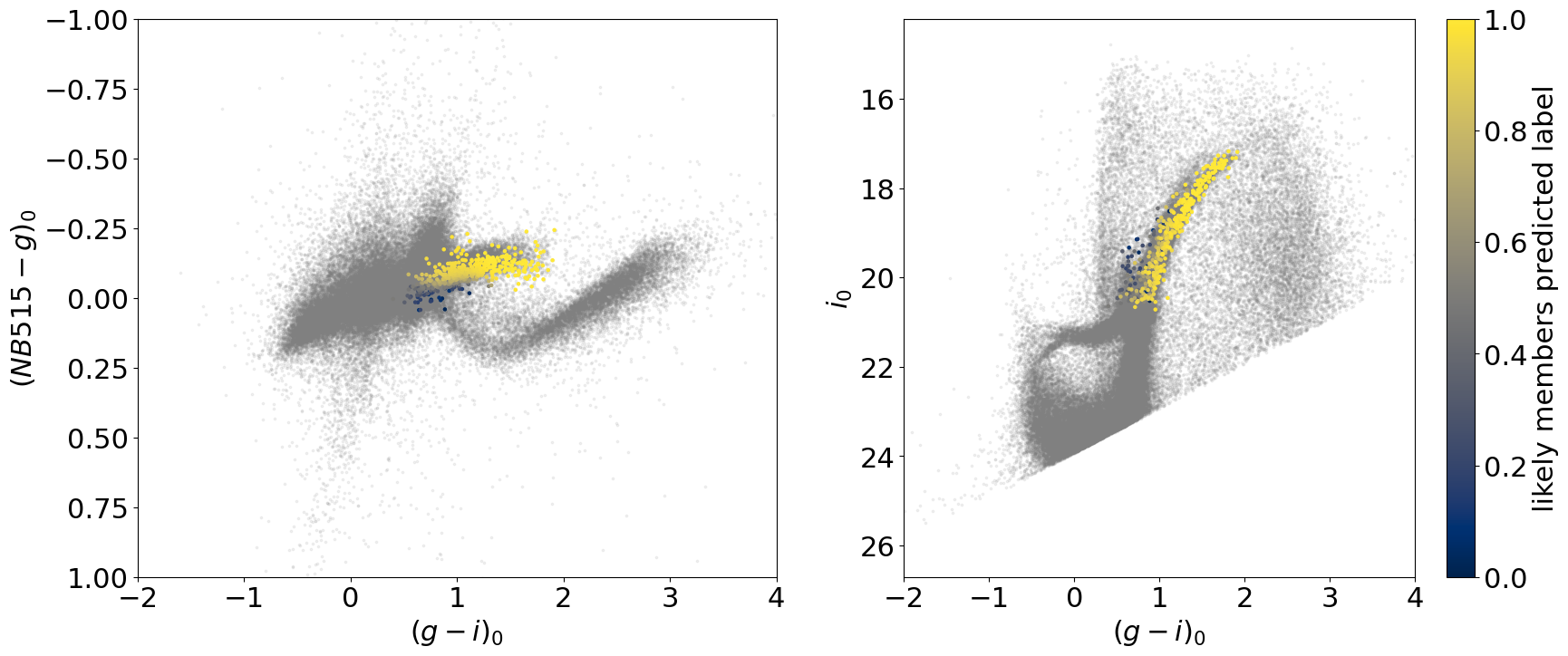}
  \caption{Two-color and color-magnitude diagrams of the Fornax Dwarf Galaxy spectroscopic members indicated by \cite{2010ApJS..191..352K}, color-coded by the predicted membership label from the machine-learning model. \label{fig:fornax_cc}}
\end{figure*}

We train the neural network with the method detailed in Section~\ref{sec:applications} to predict the \lq likely-members' label of stars in the Fornax field. Here we interpret our results as `likely members' since in addition to dwarf/giant separation, the model is informed to  include also horizontal branch stars, blue stragglers, and some dwarf stars as likely members of Fornax. As shown in 
Figure~\ref{fig:fornax}, the model can identify both the giant branch and bluer populations as likely members. Although the modeling process of member stars does not include young dwarf stars or  turnoff stars in Fornax with $-1 < (g-i)_0 < 1$, they are confidently selected as members since they are bluer than the well-defined blue edge of the Milky Way halo. To evaluate the performance, we performed a 
1~arcsecond crossmatch of the HSC Fornax catalog with the membership catalog from \cite{2010ApJS..191..352K}, which allows us to measure the accuracy quantitatively. The cross-match reveals an 85\% classification accuracy for the sample when applying a member threshold at 0.85 in the predicted label (see Figure~\ref{fig:fornax_cc}). For stars classified differently, the inconsistency arises likely due to differing approaches in defining member stars when constructing our catalog: \cite{2010ApJS..191..352K} primarily uses spectroscopic radial velocities to eliminate contaminants, while our work relies solely on narrow-band imaging to find the dwarf/giant separation, and non-members can leak through any method to assign membership. 

\subsection{M31}
As mentioned in Section \ref{sec:population}, we trained two separate machine-learning models with the same structure but different, field-specific training sets for the two selected target galaxies (Fornax and M31). This approach accounts for the variation in stellar populations, including both foreground and member stars, across these fields. In the case of M31, we interpret the prediction label as \lq likely red giants' since the model is specifically designed for dwarf/giant separation.
 \begin{figure*}[ht]
                \centering
                 \includegraphics[width=0.95\linewidth]{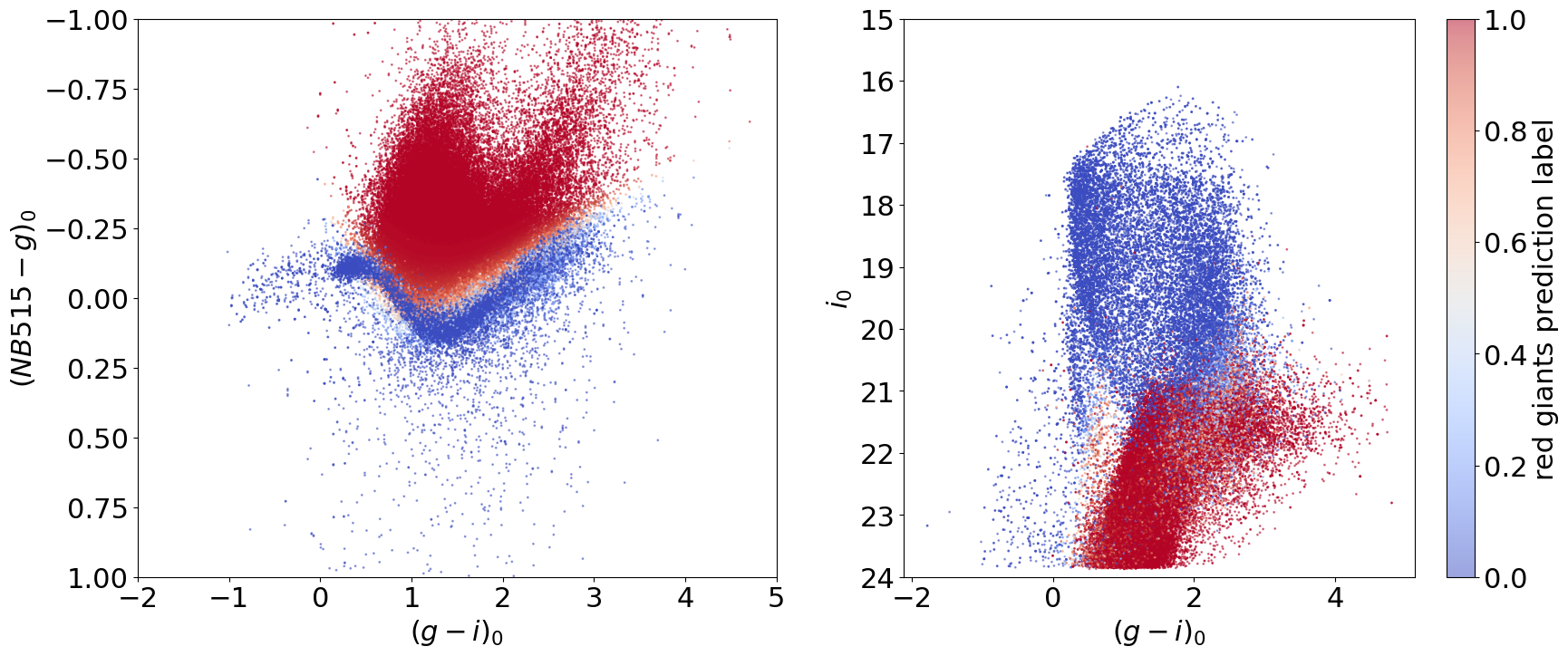}
\caption{Two-color diagram and color-magnitude diagram of a field in the inner halo of M31, color-coded by the predicted red-giant label.
\label{fig:m31_center}}
\end{figure*}

 \begin{figure*}[ht]
                \centering
                 \includegraphics[width=0.95\linewidth]{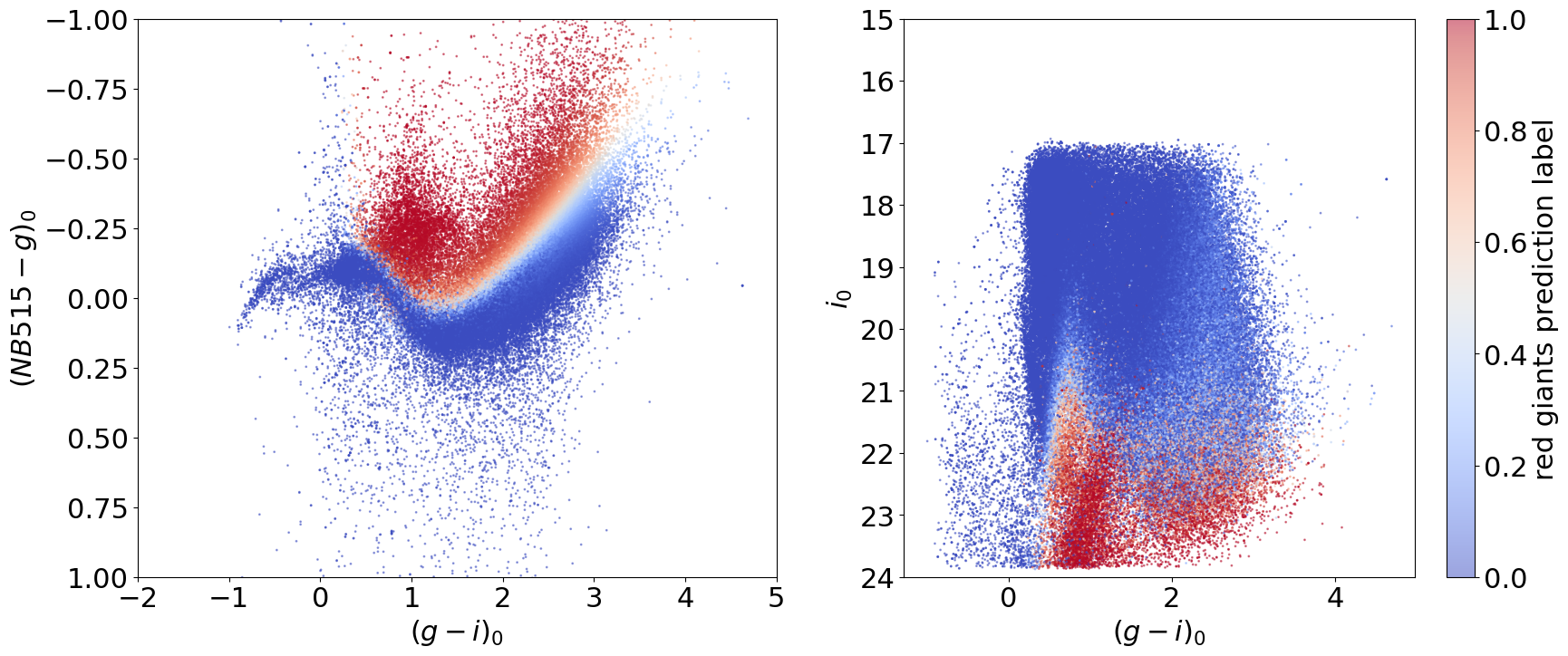}
\caption{Two-color diagram and color-magnitude diagram of a field at the NW stream of M31, color-coded by the predicted red-giant label.
\label{fig:m31_nw}}
\end{figure*}

\begin{figure*}[ht]
                \centering
                 \includegraphics[width=0.95\linewidth]{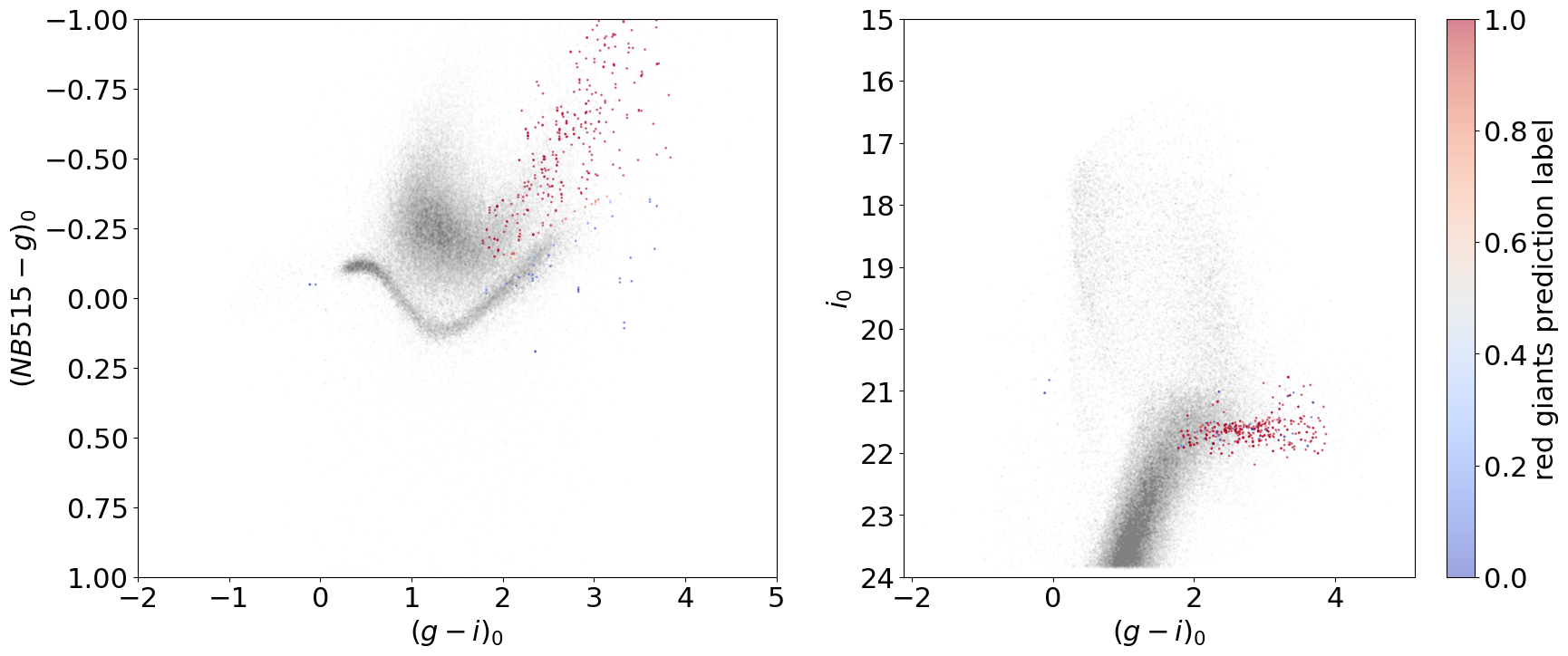}
\caption{Two-color diagram and color-magnitude diagram of a field in the inner halo of M31. The cross-matched sample with the DESI spectroscopic members is color-coded by the predicted membership label.
\label{fig:m31_desi}}
\end{figure*}

Figures \ref{fig:m31_center} and \ref{fig:m31_nw} show the two-color diagram and color-magnitude diagram for each of the two M31 fields, color-coded by the predicted membership probability. For the field in  the inner halo of M31, the model selects likely red giants that extend to the regime of extreme M-stars at $(g-i)_0=4$.  Fewer red stars with $(g-i)_0>2$ are selected in the NW field since the halo of M31 tends to be more metal-poor and less dense, and there is also a gradient in the foreground star counts with more Milky Way stars in the NW field. To evaluate the performance of our model when applied to M31, we compare our selection with the membership predictions from the DESI observations of M31 \citep{dey23}. The DESI observations primarily targeted red stars in M31, with $g-i > 2$, making the membership catalog particularly informative for evaluating our model's performance with M stars. We perform a 1 arc-second crossmatch between our M31 HSC catalog and the DESI spectroscopically selected sample of M31 sources and find 413 matched stars in the inner halo field, with no matches in the NW field. This crossmatched sample is color-coded by the predicted membership label in 
Fig.~\ref{fig:m31_desi}, while all data points of the M31 fields are plotted in gray in the background. Applying a membership threshold at 0.85 in the predicted label, our model selects 87.7\%  of the sample as likely giants. Most misclassified stars scatter at non-physical regions of the two-color diagram with large photometric errors and low precision.

\begin{figure*}[ht]
                \centering
                 \includegraphics[width=0.55\linewidth]{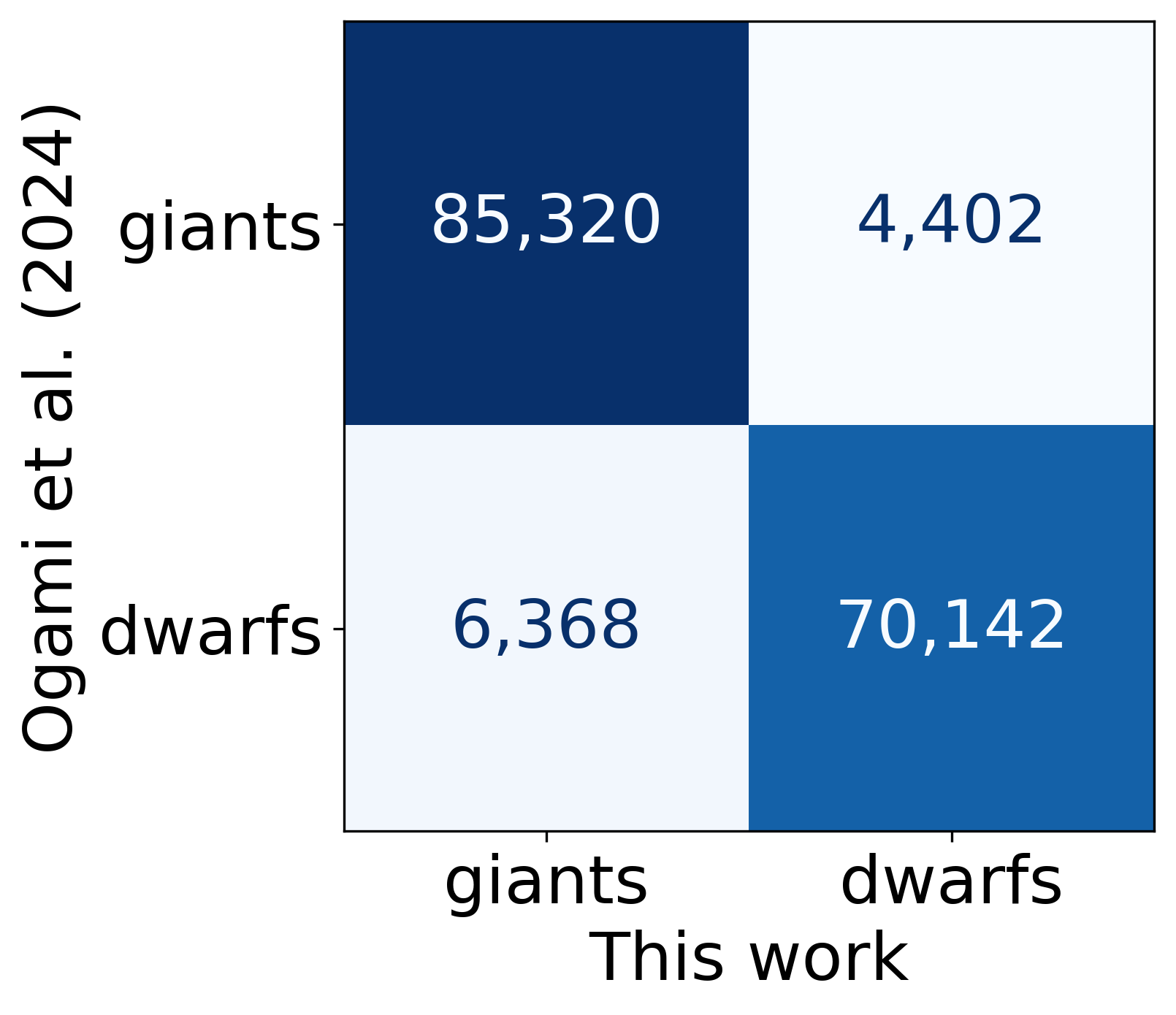}
\caption{Confusion matrix for giant/dwarf classification in M31. Predicted labels with giant probability greater than 0.85 are compared against reference labels from \cite{oga24} with $\rho_{M31}$ greater than 0.9. Diagonal elements show agreements, while off-diagonals indicate disagreements.
\label{fig:conf}}
\end{figure*}

In addition to a comparison with the spectroscopically selected DESI catalog, the fact that we utilized the  photometric data from \cite{oga24} in our study  allows for a more direct test of our results, using a larger sample size, and broader stellar population coverage. While both our method and \cite{oga24} estimate M31 membership probability, the results cannot be simply compared on the same scale, due to differences in our approaches and target functions. We adopt the definition of likely M31 RGB in \cite{oga24} with $\rho_{M31}>0.9$ (which combines both spatial information and $NB515$, as well as a color-magnitude cut) to separate likely dwarfs and red giants, and applied a threshold at 0.85 in our predicted label for comparison. We consider only stars with $(g-i)_0 < 2.5$ in both catalogs. The comparison reveals an accuracy of 93.4\% (see Figure~\ref{fig:conf}). However, visual inspection of the two-color diagrams reveals an inconsistency in the width of the dwarf sequence: our dwarf sequence is generally narrower at $1 < (g-i)_0 < 2$ and wider at $(g-i)_0 > 2$. This results in a higher mismatch rate at $1 < (g-i)_0 < 2$ but allows for the inclusion of more likely M giants in our selection.

\eject
\subsection{Limitations and Known Issues} \label{sec:limitation}
By refining our model and analyzing stars likely to be misclassified, we have identified limitations and known issues that could contribute to the misclassification rate of our machine-learning model. The primary limitation lies in our stellar-population modeling process. We make the simplified assumption that the fields of the target galaxies contain only foreground main-sequence dwarfs and member RGB/AGB stars, excluding other stellar types like blue stragglers and horizontal branch stars. This simplification arises in part due to the limited availability of spectra for these stellar types in both empirical spectral libraries and theoretical models, thus preventing us from accurately assessing their $NB515$ colors. We also oversimplify the foreground Milky Way model by adopting smooth stellar density profiles and oversimplify our target galaxies by assuming a given metallicity distribution for the member stars. Additionally, the machine-learning model is highly sensitive to the density contrast between dwarfs and giants in the training set. Modeling the density of likely member stars, particularly in M31, is challenging due to the density gradients across different fields. When constructing the training set, it is crucial to balance the trade-off between false positive and false negative predictions, especially at the red end of the color-magnitude diagram where the separation between dwarf and giant sequences is minimal.

\subsection{Advantages and Limitations of Machine Learning}
Incorporating machine learning into photometric dwarf-giant classification presents both advantages and limitations. On the positive side, machine learning offers a more objective and data-driven approach compared to traditional methods, such as manually drawing boundaries on a two-color diagram. Traditional approaches often rely on heuristic or visually defined separation criteria, which are inherently susceptible to subjective interpretation and observer-dependent biases. In contrast, ML algorithms learn decision boundaries directly from the data through optimized statistical processes, minimizing human intervention and biases. In addition, the machine learning method can be generalized into new datasets with minimal changes needed. Another key strength of machine learning is its ability to process multiple features simultaneously in high-dimensional space, enabling the use of the full information in classification compared to traditional two-dimensional cuts. This multivariate analysis enhances discriminative power, particularly when stellar populations exhibit overlapping features in color or magnitude space.

Despite the success of machine learning in astrophysical applications, certain limitations persist due to the inherent characteristics of these algorithms. Although neural networks are well-established as universal approximations \citep{HORNIK1989359} and perform effectively in tasks such as classification and regression, their black-box nature introduces challenges in interpretability and explainability. This lack of interpretability complicates efforts to identify potential biases or understand the underlying decision-making process of the model. For example, in our classification framework, while the model outputs a prediction score indicating the likelihood of a star being a red giant, these scores should not be interpreted as calibrated statistical probabilities. Specifically, a prediction score of 0.9 suggests a higher probability of being a red giant compared to a score of 0.1; however, the raw outputs are not normalized to a physically meaningful scale. Consequently, a direct comparison between intermediate scores, such as concluding that a star with a score of 0.6 is more likely to be a red giant than one with 0.4, may not be statistically justified without proper calibration. This limitation underscores the need for careful interpretation of ML-derived predictions. 

Furthermore, the machine learning process itself relies on empirically chosen hyperparameters that were selected after testing a variety of options, including model architecture, learning rate, batch size, and number of training epochs, rather than being guided by physical principles. This empirical nature introduces an additional layer of uncertainty, as different hyperparameter choices may lead to varying model behaviors without a clear theoretical justification.

\section{Conclusion} \label{sec:conclusion}

The $NB515$ narrowband filter has long been recognized as a valuable tool for distinguishing dwarf and giant stars, particularly in metal-poor populations, thanks to its sensitivity to surface gravity features. However, its application to cooler, more metal-rich stars, particularly M-stars, has remained underexplored. In this work, we have demonstrated that, when combined with machine learning techniques, $NB515$ photometry can effectively classify stellar populations across a wider range of metallicities and temperatures, including the challenging M-star regime.

Using synthetic photometry derived from the high-quality, flux-calibrated MaStar and X-Shooter spectral libraries, we systematically modeled the response of the $NB515$ filter to key stellar atmospheric parameters, including effective temperature, metallicity, and surface gravity. This allowed us to identify the specific photometric signatures that differentiate metal-rich giants from foreground dwarfs, even in the presence of degeneracies that complicate traditional color-based selection methods. Building on these models, we developed a neural network-based classifier that leverages $NB515$ photometry alongside broad-band filters to achieve robust stellar classification without requiring spectroscopic pre-selection. Our machine learning framework was rigorously tested on M31 and the Fornax dwarf spheroidal galaxy, where it successfully distinguished extragalactic red giants from Milky Way dwarf contaminants with an accuracy exceeding 85\% when validated against existing spectroscopic and photometric membership catalogs. Notably, the method performs well even for the coolest M-stars, where molecular absorption bands and metallicity effects complicate traditional photometric classification. This represents a significant expansion of $NB515$'s utility, as prior applications were largely confined to warmer, more metal-poor stars.

The success of this approach has important implications for stellar population studies in the Local Group. By enabling efficient dwarf/giant separation for metal-rich M-stars, a population that is abundant in the central region of M31 and other disk galaxies but difficult to characterize, our method opens new avenues for studying star formation histories, chemical enrichment, and dynamical evolution in systems like M31 and the Milky Way satellites. Furthermore, the technique is particularly valuable for optimizing spectroscopic follow-up campaigns, as it allows for pre-observation target screening to minimize contamination and maximize observing efficiency.

\section{acknowledgments}
We thank the referee for their constructive feedback. KD, CF, RFGW, LD, and ASS acknowledge support through the generosity of Eric and Wendy Schmidt, by recommendation of
the Schmidt Futures program.  RFGW
thanks her sister, Katherine Barber, for her support.

Funding for the Sloan Digital Sky 
Survey IV has been provided by the 
Alfred P. Sloan Foundation, the U.S. 
Department of Energy Office of 
Science, and the Participating 
Institutions. 

SDSS-IV acknowledges support and 
resources from the Center for High 
Performance Computing  at the 
University of Utah. The SDSS 
website is www.sdss4.org.

SDSS-IV is managed by the 
Astrophysical Research Consortium 
for the Participating Institutions 
of the SDSS Collaboration including 
the Brazilian Participation Group, 
the Carnegie Institution for Science, 
Carnegie Mellon University, Center for 
Astrophysics | Harvard \& 
Smithsonian, the Chilean Participation 
Group, the French Participation Group, 
Instituto de Astrof\'isica de 
Canarias, The Johns Hopkins 
University, Kavli Institute for the 
Physics and Mathematics of the 
Universe (IPMU) / University of 
Tokyo, the Korean Participation Group, 
Lawrence Berkeley National Laboratory, 
Leibniz Institut f\"ur Astrophysik 
Potsdam (AIP),  Max-Planck-Institut 
f\"ur Astronomie (MPIA Heidelberg), 
Max-Planck-Institut f\"ur 
Astrophysik (MPA Garching), 
Max-Planck-Institut f\"ur 
Extraterrestrische Physik (MPE), 
National Astronomical Observatories of 
China, New Mexico State University, 
New York University, University of 
Notre Dame, Observat\'ario 
Nacional / MCTI, The Ohio State 
University, Pennsylvania State 
University, Shanghai 
Astronomical Observatory, United 
Kingdom Participation Group, 
Universidad Nacional Aut\'onoma 
de M\'exico, University of Arizona, 
University of Colorado Boulder, 
University of Oxford, University of 
Portsmouth, University of Utah, 
University of Virginia, University 
of Washington, University of 
Wisconsin, Vanderbilt University, 
and Yale University.

This work has made use of data from the European Space Agency (ESA) mission
{\it Gaia} (\url{https://www.cosmos.esa.int/gaia}), processed by the {\it Gaia}
Data Processing and Analysis Consortium (DPAC,
\url{https://www.cosmos.esa.int/web/gaia/dpac/consortium}). Funding for the DPAC
has been provided by national institutions, in particular the institutions
participating in the {\it Gaia} Multilateral Agreement.

Based on data from the The X-Shooter Spectral Library service developed by the Spanish Virtual Observatory in the framework of the IAU Comission G5 Working Group : Spectral Stellar Libraries

The national facility capability for SkyMapper has been funded through ARC LIEF grant LE130100104 from the Australian Research Council, awarded to the University of Sydney, the Australian National University, Swinburne University of Technology, the University of Queensland, the University of Western Australia, the University of Melbourne, Curtin University of Technology, Monash University and the Australian Astronomical Observatory. SkyMapper is owned and operated by The Australian National University's Research School of Astronomy and Astrophysics. The survey data were processed and provided by the SkyMapper Team at ANU. The SkyMapper node of the All-Sky Virtual Observatory (ASVO) is hosted at the National Computational Infrastructure (NCI). Development and support of the SkyMapper node of the ASVO has been funded in part by Astronomy Australia Limited (AAL) and the Australian Government through the Commonwealth's Education Investment Fund (EIF) and National Collaborative Research Infrastructure Strategy (NCRIS), particularly the National eResearch Collaboration Tools and Resources (NeCTAR) and the Australian National Data Service Projects (ANDS).

The Pan-STARRS1 Surveys (PS1) and the PS1 public science archive have been made possible through contributions by the Institute for Astronomy, the University of Hawaii, the Pan-STARRS Project Office, the Max-Planck Society and its participating institutes, the Max Planck Institute for Astronomy, Heidelberg and the Max Planck Institute for Extraterrestrial Physics, Garching, The Johns Hopkins University, Durham University, the University of Edinburgh, the Queen's University Belfast, the Harvard-Smithsonian Center for Astrophysics, the Las Cumbres Observatory Global Telescope Network Incorporated, the National Central University of Taiwan, the Space Telescope Science Institute, the National Aeronautics and Space Administration under Grant No. NNX08AR22G issued through the Planetary Science Division of the NASA Science Mission Directorate, the National Science Foundation Grant No. AST–1238877, the University of Maryland, Eotvos Lorand University (ELTE), the Los Alamos National Laboratory, and the Gordon and Betty Moore Foundation.

The Hyper Suprime-Cam (HSC) collaboration includes the astronomical communities of Japan and Taiwan, and Princeton University. The HSC instrumentation and software were developed by the National Astronomical Observatory of Japan (NAOJ), the Kavli Institute for the Physics and Mathematics of the Universe (Kavli IPMU), the University of Tokyo, the High Energy Accelerator Research Organization (KEK), the Academia Sinica Institute for Astronomy and Astrophysics in Taiwan (ASIAA), and Princeton University. Funding was contributed by the FIRST program from Japanese Cabinet Office, the Ministry of Education, Culture, Sports, Science and Technology (MEXT), the Japan Society for the Promotion of Science (JSPS), Japan Science and Technology Agency (JST), the Toray Science Foundation, NAOJ, Kavli IPMU, KEK, ASIAA, and Princeton University. 

This paper makes use of software developed for the Large Synoptic Survey Telescope. We thank the LSST Project for making their code available as free software at  http://dm.lsst.org

Based [in part] on data collected at the Subaru Telescope and retrieved from the HSC data archive system, which is operated by Subaru Telescope and Astronomy Data Center at National Astronomical Observatory of Japan.

We acknowledge that this research makes use of SciServer, a resource developed and operated by the Johns Hopkins University, Institute for Data Intensive Engineering and Science (IDIES).

\software{\texttt{NumPy} \citep{numpy}, \texttt{SciPy} \citep{scipy}, \texttt{astropy} \citep{astropy_2013, astropy_2018, astropy_2022}, \texttt{Matplotlib} \citep{matplotlib}, \texttt{TOPCAT} \citep{tay11}, \texttt{PyTorch}, \citep{pytorch}}

\eject
\appendix
\renewcommand{\thefigure}{\Alph{section}.\arabic{figure}} 
\setcounter{figure}{0} %

\section{$NB515$'s Response to [$\alpha$/Fe]} \label{app:alpha}

\begin{figure*}[ht]
                \centering
                 \includegraphics[width=0.95\linewidth]{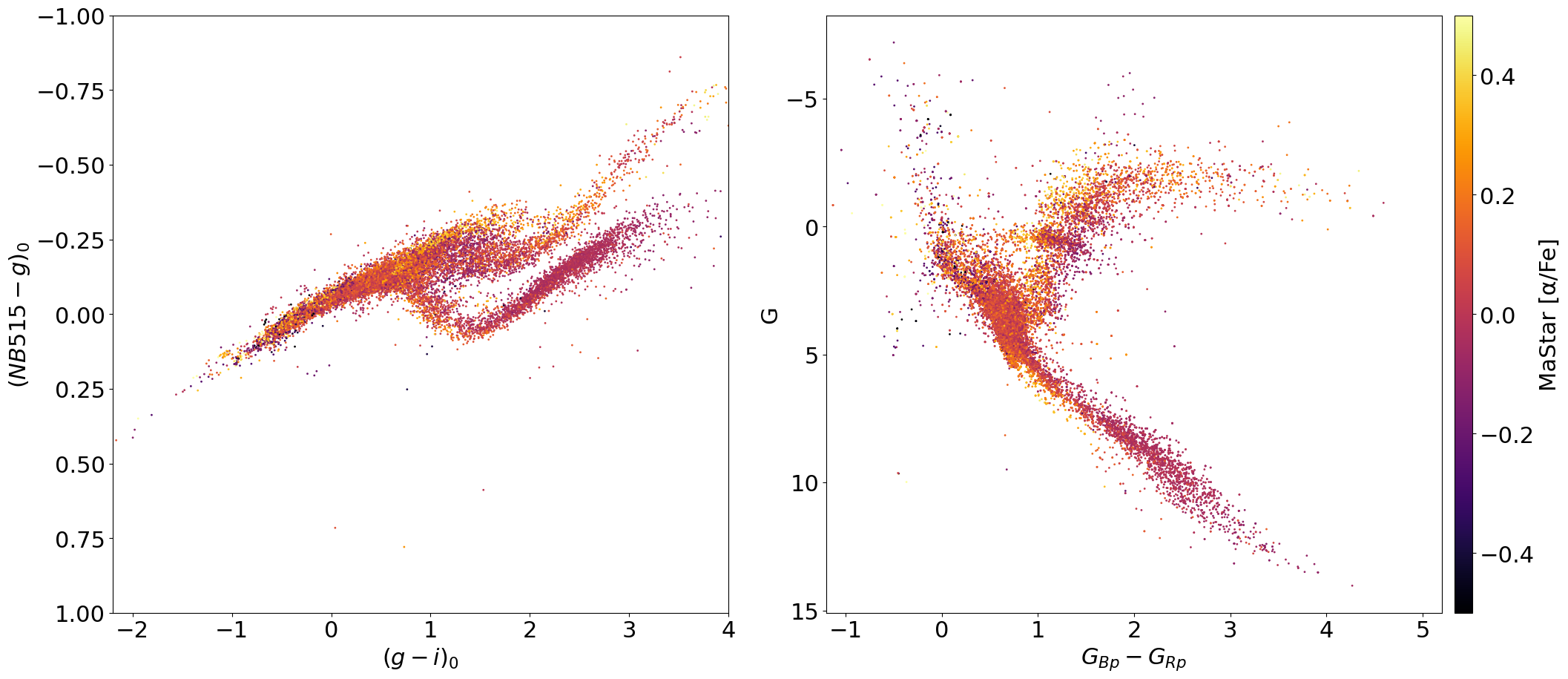}
\caption{$Gaia$ color-magnitude diagram of stars in the MaStar Library and two-color diagram of MaStar synthetic photometry, color-coded by [$\alpha$/Fe] from the mean MaStar Library parameters.
\label{fig:alpha}}
\end{figure*}

Since the $NB515$ filter is centered on the MgH+Mgb absorption features at 515 nm, the $(NB515-g)_0$ color is also sensitive to the stellar alpha-elements and magnesium abundances. We present the $Gaia$ color-magnitude diagram of stars in the MaStar Library and two-color diagram of MaStar synthetic photometry, color-coded by [$\alpha$/Fe] from the mean MaStar Library parameters in Figure \ref{fig:alpha}. A more detailed analysis of $NB515$ filter's sensitivity to chemical abundances is analyzed by \cite{hong25}.

\setcounter{figure}{0} %

\section{Comparison with DDO51} \label{app:ddo51}
To further understand how $NB515$-based synthetic colors compare to DDO51 colors, we conducted a star-by-star comparison between the $NB515$ magnitudes and the $DDO51$ magnitudes using the cross-match of our MaStar synthetic photometry catalog with the APOGEE stellar parameters catalog \citep{2022ApJS..259...35A}. $DDO51$ is an intermediate-band filter observing the same MgH+Mgb absorption features at 515 nm \citep{cla79}, and \cite{maj00} proposed a technique to photometrically separate RGB stars and dwarfs based on the ($M-DDO51$) color index, where $M$ is a broadband filter in the Washington system. The $DDO51$ filter has been employed in the APOGEE target selection to select giant stars in the Milky Way halo and explore the halo substructures. 

 \begin{figure*}[ht]
                \centering
                 \includegraphics[width=0.95\linewidth]{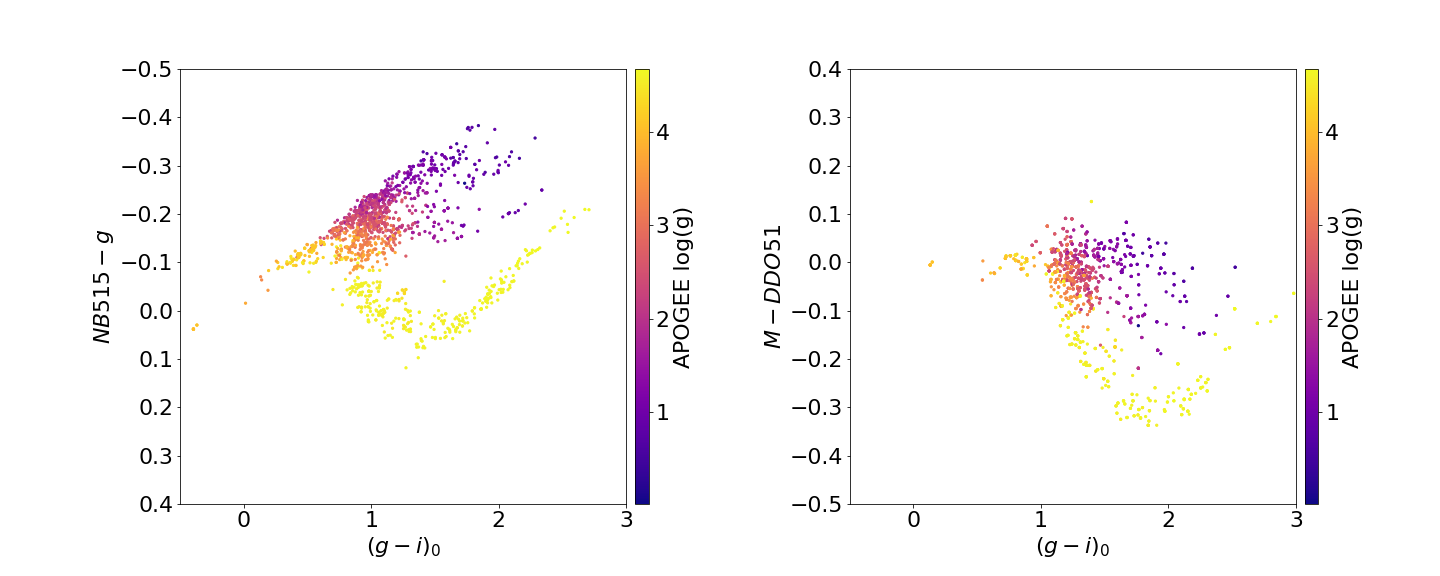}
\caption{The $M-DDO51$ and $NB515-g$ two-color diagrams of the crossmatched sample between our MaStar synthetic photometry catalog and the APOGEE stellar parameters catalog; both use synthetic $(g-i)_0$ along the $x$-axis for ease of comparison. The diagrams exhibit similar dwarf-giant separation, despite the different shapes of the giant sequences.
\label{fig:ddo51}}
\end{figure*}

Figure \ref{fig:ddo51} presents the $(g-i)_0 - (M-DDO51)$ and $(g-i)_0 - (NB515-g)_0$ two-color diagrams of the cross-matched sample between our derived MaStar synthetic photometry catalog with the APOGEE stellar parameters catalog, color-coded by $\log(g)$ from the APOGEE Stellar Parameters and Abundances Pipeline \citep[ASPCAP,][]{2018AJ....156..125H}. While \cite{2000AJ....120.2550M} used $M-T_2$ color index as the effective temperature indicator to separate the dwarfs and giants on the two-color diagram, we opted for $(g-i)_0$ to ensure comparability along the $x$-axis. Both two-color diagrams effectively separate dwarfs and giants at a color range $1< (g-i)_0 < 2.5$. Although the shapes of the dwarf sequences on the two diagrams are similar, the giant sequence exhibits a broader spread along the M-DDO51 color index compared to $NB515 - g$.  This disparity likely comes from the width difference between DDO51 and $NB515$, with $NB515$ being narrower and thus more sensitive to the Mg line depth at 515 nm.

\setcounter{figure}{0} %

\section{Crossmatching with Photometry Surveys}\label{app:cm}
 \begin{figure*}[ht]
                \centering
                 \includegraphics[width=0.95\linewidth]{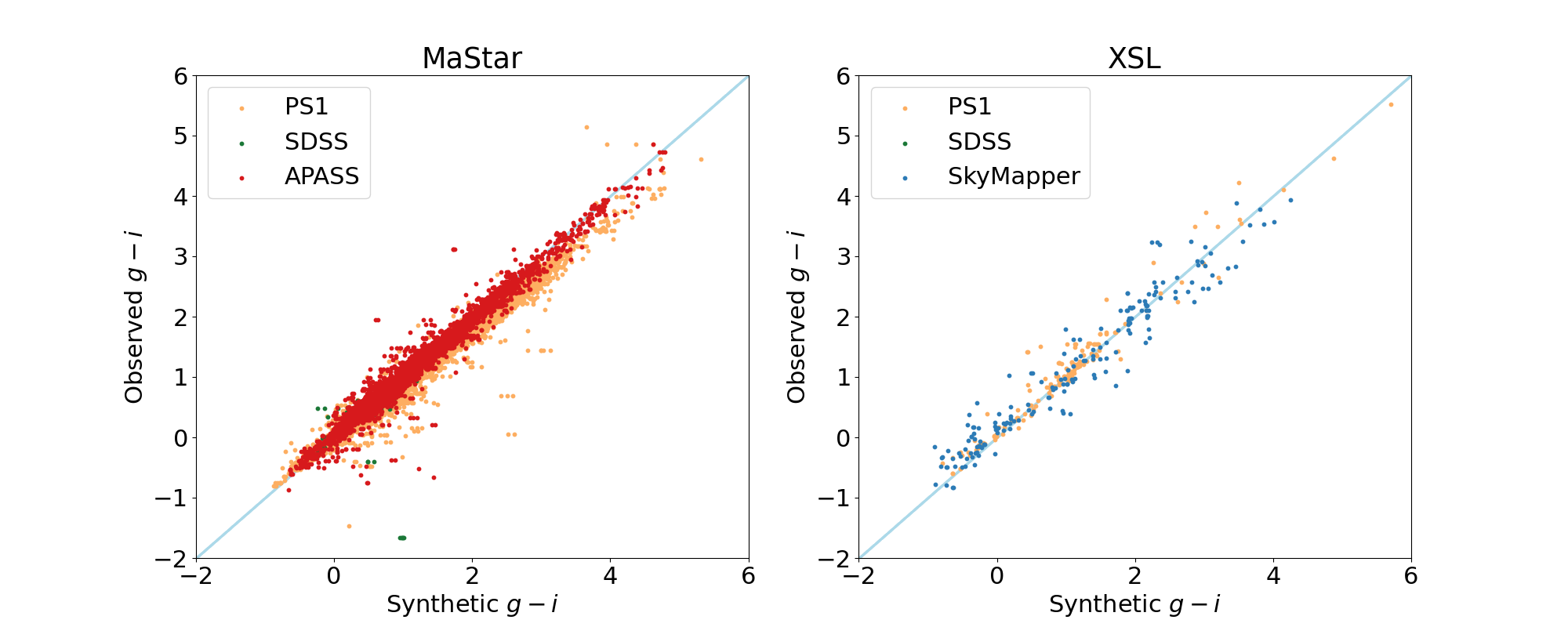}
\caption{Comparison of synthetic HSC $g-i$ colors with observed $g-i$ colors.
\label{fig:obs_color}}
\end{figure*}

To assess the accuracy of our synthetic color calculations, we crossmatch our synthetic color catalog with photometric surveys such as SDSS, Pan-STARRS 1, APASS, and SkyMapper, as described in Section \ref{sec:data}. 
The MaStar catalog has 16589 matches with Pan-STARRS 1, 3169 matches with SDSS, and 6628 matches with APASS, and the XSL catalog has 174 matches with SDSS, 412 matches with Pan-STARRS 1, and 465 matches with SkyMapper. After excluding stars with mismatches and retaining the \lq best-match' value for stars with multiple matches, we finalized a crossmatch dataset comprising 11207 entries from MaStar and 337 entries from XSL. We then perform a one-to-one comparison between the synthetic and observed $g-i$ colors. The comparison reveals a generally consistent trend, with a Root Mean Square Error (RMSE) of 0.15. A slight discrepancy appears, particularly at $g-i>2$, likely due to minor differences in filter design across these surveys.

\section{Neural Network Architecture and Training Method}\label{app:ml}
Our model is a neural network composed of 5 linear layers (also called fully connected layers). Each layer is a set of weighted connections between input features --- such as $(g-i)_0$ --- and the next layer of neurons (nodes) in the network. The weights are adjusted during training to better predict membership probabilities.

Between each of these layers, we apply a Sigmoid activation function. The Sigmoid function (Eq.\ref{eq:sig}) is a bounded, differentiable, \lq S-shaped' function that introduces non-linearity to the decision boundaries:
\begin{equation}
    S(x) = \frac{1}{1+e^{-x}}
    \label{eq:sig}
\end{equation}
 and thus the boundaries will be curves rather than straight lines.

At the final layer of the network, we use a Softmax function (Eq.~\ref{eq:soft}) to transform the model's raw output into labels similar to probabilities. The Softmax function takes a vector of real numbers (the model's unnormalized predictions) and converts it into a normalized distribution. In our case, it ensures that the two outputs, $P_{member}$ and $P_{nonmember}$, add up to 1, making them interpretable similar to probabilities.
\begin{equation}
    p(\hat{z})_i = \frac{e^{z_i}}{\sum_{j=1}^K e^{z_j}}
    \label{eq:soft}
\end{equation}

We use a Binary Cross-entropy loss function (Eq.~\ref{eq:bce}) to evaluate how well the model performs during training. The loss function measures the difference between the predicted probability vector $p(\hat{z})$ (what the model outputs) and the true label vector $q(\hat{z})$ (what the correct answer is). The Binary Cross-entropy loss is measured as a number between 0 and 1, with 0 being a perfect model and the machine-learning model aims to minimize the loss.

\begin{equation}
    H(P, Q) = - \sum_x p(\hat{z})log(q(\hat{z}))
    \label{eq:bce}
\end{equation}

To minimize the loss, we use an optimization algorithm called RMSprop (Root Mean Square Propagation, a variant of Stochastic Gradient Descent, or SGD). The basic idea of SGD is to adjust the model's parameters iteratively (like the weights in a neural network) in small steps, moving in the direction that reduces the loss. The size of each step is controlled by a parameter called the learning rate. By repeatedly adjusting the parameters in this way, the model gradually improves its predictions until it finds the set of parameters that minimizes the loss. RMSprop improves upon SGD by using an adaptive learning rate. It adjusts the learning rate based on how fast the loss changes in different directions, which helps the model converge faster and avoid overshooting the optimal solution.

\end{CJK*}

\bibliography{ref}{}
\bibliographystyle{aasjournal}
\end{document}